\documentclass[aps,prl,twocolumn,reprint,superscriptaddress]{revtex4}
\usepackage{graphicx}
\usepackage{amsmath}
\usepackage{amssymb}
\usepackage{colordvi}
\usepackage{mathrsfs}
\usepackage{bm}
\usepackage{verbatim}
\usepackage{dcolumn}
\usepackage{epsfig}
\usepackage{subfigure}

\begin{document}
    \title{Large-Gap Quantum Anomalous Hall Effect in Monolayer Halide Perovskite}
    \author{Zeyu Li}
	\affiliation{ICQD, Hefei National Laboratory for Physical Sciences at Microscale, CAS Key Laboratory of Strongly-Coupled Quantum Matter Physics, and Department of Physics, University of Science and Technology of China, Hefei, Anhui 230026, China}
	\author{Yulei Han}
	\affiliation{ICQD, Hefei National Laboratory for Physical Sciences at Microscale, CAS Key Laboratory of Strongly-Coupled Quantum Matter Physics, and Department of Physics, University of Science and Technology of China, Hefei, Anhui 230026, China}
	\author{Zhenhua Qiao}
	\email[Correspondence author:~]{qiao@ustc.edu.cn}
\affiliation{ICQD, Hefei National Laboratory for Physical Sciences at Microscale, CAS Key Laboratory of Strongly-Coupled Quantum Matter Physics, and Department of Physics, University of Science and Technology of China, Hefei, Anhui 230026, China}
	\date{\today}
	\begin{abstract}
      We theoretically propose a family of structurally stable monolayer halide perovskite A$_3$B$_2$C$_9$ (A=Rb, Cs; B=Pd, Pt; C=Cl, Br) with easy magnetization planes. These materials are all half-metals with large spin gaps over 1~eV accompanying with a single spin Dirac point located at K point. When the spin-orbit coupling is switched on, we further show that Rb$_3$Pt$_2$Cl$_9$,  Cs$_3$Pd$_2$Cl$_9$, and Cs$_3$Pt$_2$Cl$_9$ monolayers can open up large band gaps from 63 to 103 meV to harbor quantum anomalous Hall effect with Chern numbers of $\mathcal{C}=\pm1$, whenever the mirror symmetry is broken by the in-plane magnetization. The corresponding Berezinskii-Kosterlitz-Thouless transition temperatures are over 248~K. Our findings provide a potentially realizable platform to explore quantum anomalous Hall effect and spintronics at high temperatures.
    \end{abstract}

\maketitle

\textit{Introduction---.} Due to the topological protection from spatial separation, the Landau-level induced chiral edge modes make quantum Hall effect appealing in designing future low-power electronics~\cite{QHE}. In early time, the requirement of strong magnetic field severely prevents the possible application of quantum Hall effect. Ever since the experimental discovery of graphene and topological insulators, various recipes were proposed to produce the quantum Hall effect without external magnetic field, i.e., quantum anomalous Hall effect (QAHE). Two representative categories are monolayer atomic crystals (e.g., graphene and graphene-like materials)~\cite{QAHEHaldane,Rashba-Graphene,Qiao-antiferromagnet} and magnetic topological insulators~\cite{Cr-Bi2Se3,Qi-Qiao-codoping,In-plane-Zhong,MBT-ST}. On one side, in 2010 it was theoretically reported that graphene can open up a band gap to harbor QAHE in the presence of both Rashba spin-orbit coupling and exchange field~\cite{Rashba-Graphene}. However, although there has been tremendous progress towards approaching the quantization of anomalous Hall effect, i.e. $1/4e^2/h$, experimental realization of QAHE in graphene is still a challenge due to the weak Rashba spin-orbit coupling~\cite{graphene-YIG-1,graphene-YIG-2}. On the other side, nearly at the same time, magnetic topological insulator was theoretically proposed to host QAHE via Cr dopants; And such a scheme was soon successfully realized in experiment at the extremely low temperature of 30 mK after three years' efforts~\cite{Exp-Cr-Bi2Se3}; When more attention was focused on how to enhance the QAHE observation temperature, it was experimentally reported that the antiferromagnetic MnBi$_2$Te$_4$ thin flakes can reach a record-breaking temperature of 6.5K with the aid of external magnetic field~\cite{Theory1-MnBiTe,Theory2-MnBiTe,QAHE-MnBiTe,P-doped-MnBiTe}.

Nevertheless, the current achievement is still further away from the practical application. As an important material family, the perovskite ABO$_3$ consists of transition metal and heavy elements~\cite{Perovskite-1,Perovskite-2,Perovskite-3}, implying a wonderland for breeding topological states, such as topological insulator, topological node-line semimetal and topological crystalline metal~\cite{Perovskite-TI1,Perovskite-TI2,Perovskite-NL,Perovskite-TCM}. Besides, several theoretical proposals have been suggested by building superlattices with a (111) bilayer rare-earth perovskite embedded inside wide-gap perovskites~\cite{Xiao-Scheme,Double-QAHE,LaNiO3,LaPdO3-guo}, where the major experimental difficulty is the requirement of precise alignment at the interfaces. Inspired by the broad interest of two-dimensional materials, the two-dimensional layered perovskites have also been experimentally prepared, e.g. 111-type In-based halide perovskite Cs$_3$In$_2$Cl$_9$~\cite{Cs3Sb2Br9-1,Cs3Sb2Br9-2}. The layered perovskite materials display the possibility of realizing the QAHE due to their large intrinsic spin-orbit coupling and magnetism~\cite{PdCl,OsCl3}.

In this Letter, we theoretically predict a new two-dimensional halide perovskite material family A$_3$B$_2$C$_9$, where A, B and C represent alkali metal element (Rb, Cs), transition metal element (Pd, Pt) and halogen element (Cl, Br), respectively. They share the same structure as the monolayer Cs$_3$In$_2$Cl$_9$ with the transition metal atoms forming buckled honeycomb lattice linked by halogen atoms. By performing first-principles calculations, we find that they are all half-metals, with a huge spin-splitting over 1.0 eV. The single spin Dirac point emerging at K point is topologically protected by the trigonal crystal symmetry. After the spin-orbit coupling is switched on, topologically nontrivial band gaps hosting QAHE open in monolayer Rb$_3$Pt$_2$Cl$_9$, Cs$_3$Pt$_2$Cl$_9$ and Cs$_3$Pd$_2$Cl$_9$, when the vertical mirror symmetry is broken by the magnetization. For the magnetization lying inside the plane, the nontrivial gaps can reach up to 103, 69 and 93 meV at the HSE06 level, greatly exceeding room temperature. Tuning the magnetization direction from in-plane to out-of-plane will raise the gaps to 137, 99 and 100 meV, respectively. Monte Carlo calculations show that these three materials own high Berezinskii-Kosterlitz-Thouless transition temperature above 248K, implying the quasi-long-range order can be observed at high temperatures like CoPS$_3$~\cite{CoPS3}. The successful synthesis of Cs$_3$Fe$_2$Cl$_9$~\cite{Cs3Fe2Cl9} strongly indicates the possibility of realizing high temperature QAHE in our proposed monolayer halide perovskites.

\begin{figure}
  \centering
  \includegraphics[width=0.45\textwidth]{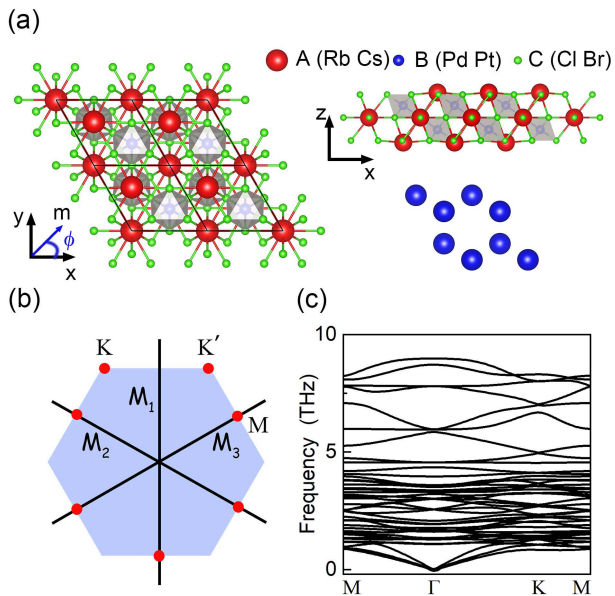}
  \caption{(a) Top view and side view of monolayer A$_3$B$_2$C$_9$, the transition metal atoms form a buckled honeycomb lattice. (b) First Brillouin zone with high symmetry points, the solid lines representing three vertical mirror planes. (c) Phonon spectra of Cs$_3$Pt$_2$Cl$_9$.}
  \label{fig1}
\end{figure}

\textit{Calculation Methods---.} Our calculations were performed using the projected-augmented-wave method as implemented in the VASP package~\cite{vasp}, and the generalized gradient approximation exchange-correlation~\cite{GGA} potential was used. The kinetic cutoff energy of plane wave was set to be 500 eV. Brillouin zone was sampled with a Gamma-centered 13$\times$13$\times$1 grid based on the scheme proposed by Monkhorst-Pack for calculation of structural optimization and magnetic property, whereas 6$\times$6$\times$1 grid was used for electronic structures~\cite{MP}. A vacuum buffer space over 20~{\AA} was included to prevent interaction between adjacent slabs. The convergence criterion was set to be 10$^{-8}$ eV and 10$^{-7}$ eV for energy in optimization and self-consistent field calculations, respectively. During structural optimization, all atoms were fully relaxed, and forces were converged to less than 0.0005 eV/{\AA}. The phonon calculations were carried out by using the density functional perturbation theory as implemented in the PHONOPY package with 3$\times$3$\times$1 supercell~\cite{phonon}. The screened hybrid functional (HSE06) method was applied to predict band gap closing to experiments~\cite{hse06}. The maximally localized Wannier functions were constructed by using the software package WANNIER90 with interfacing VASP~\cite{wannier90,wannier90-QAHE}. The anomalous Hall conductivity was obtained by the interpolation of maximally localized Wannier functions. The anomalous Hall conductivity was obtained by summing Berry curvatures over all occupied valence bands:
\begin{equation}
\sigma_{\alpha\beta}=-\frac{e^2}{\hbar}\int_{BZ}\frac{dk}{(2\pi)^3}\begin{matrix}\sum_{n}\end{matrix}f_{n}(k)\Omega_{n,\alpha\beta}(k)
\end{equation}
The edge state was calculated by iterative Green's function method as implemented in the WannierTools package~\cite{wannier_tools}. Mcsolver code was used to perform Monte Carlo simulations on a 30$\times$30 supercell with 200,000 steps at each temperature~\cite{Monte}.

\begin{table}
	\centering
	\renewcommand\arraystretch{2}
	\begin{tabular}{c|c|c|c|c|c|c}
		\hline\hline
		~ & a & c & ~ & J$_1$ & J$_2$ & J$_3$ \\
		~ & \r{A} & \r{A} & GS & (meV) & (meV) & (meV)\\\hline
		Rb$_3$Pd$_2$Cl$_9$ & 7.036 & 5.763 & FM & -136.0 & -1.7 & 4.2\\
		Rb$_3$Pd$_2$Br$_9$ & 7.349 & 5.919 & FM & -9.0 & -2.0 & -3.0\\
		Rb$_3$Pt$_2$Cl$_9$ & 7.061 & 5.756 & FM & -123.6 & -6.1 & 8.9\\
		Rb$_3$Pt$_2$Br$_9$ & 7.385 & 5.898 & FM & -20.5 & -7.8 & -3.3\\
		Cs$_3$Pd$_2$Cl$_9$ & 7.215 & 6.091 & FM & -173.3 & 2.3 & 9.1\\
		Cs$_3$Pd$_2$Br$_9$ & 7.512 & 6.252 & FM & -45.6 & -4.1 & -1.7\\
		Cs$_3$Pt$_2$Cl$_9$ & 7.209 & 6.098 & FM & -115.8 & -20.4 & 23.8\\
		Cs$_3$Pt$_2$Br$_9$ & 7.582 & 6.203 & FM & -59.2 & -9.9 & 1.1\\
		\hline\hline
	\end{tabular}
	\label{table1}
	\caption{Lattice constants,  magnetic ground state (GS) and exchange coupling constants of monolayer A$_3$B$_2$C$_9$.}
\end{table}

\textit{Structural and Magnetic Properties---.} Monolayer A$_3$B$_2$C$_9$ is crystallized in P-3m1 space group (No. 164), as depicted in Fig.~\ref{fig1}(a). There are three vertical mirror planes across high symmetry M point in the corresponding first Brillouin zone as shown in Fig.~\ref{fig1}(b). Its primitive cell contains three alkaline metal atoms, two transition metal atoms and nine halogen atoms. Each transition metal atom is surrounded by six nearest halogen atoms forming distorted octahedra [see Fig. S1]. Monolayer A$_3$B$_2$C$_9$ exhibits five atomic layers, with transition metal atoms being located at different layers to form a buckled honeycomb lattice in favor of acquiring Dirac point in electronic structure. The interaction between two transition metal atoms is bridged by a halogen atom forming the superexchange mechanism. Similar structure can be found in (111) bilayer transition metal oxides perovskite~\cite{LaNiO3}.
The alkali metal and halogen atoms at the outermost layers form a shield to protect the inner buckled honeycomb lattice composed of transition metal atoms from degradation of environment in monolayer A$_3$B$_2$C$_9$, leading to high stability of the structures as experimentally synthesized halide perovskites Cs$_3$In$_2$Cl$_9$ and Cs$_2$PtI$_6$~\cite{Cs3Sb2Br9-2,Cs2PtI6}. The structural stability of monolayer A$_3$B$_2$C$_9$ is also confirmed by phonon spectrum calculations [see Fig.~\ref{fig1}(c) and Fig. S2]. The optimized lattice constants of monolayer A$_3$B$_2$C$_9$ (A=Rb, Cs; B=Pd, Pt; C=Cl, Br) are listed in Table~I. Along with the increase of atomic number of component elements, the lattice constants increase monotonically, ranging from 7.036 to 7.582 {\AA} within $x$-$y$ plane, and 5.763 to 6.252 {\AA} along z direction.

\begin{figure}
  \centering
  \includegraphics[width=0.45\textwidth]{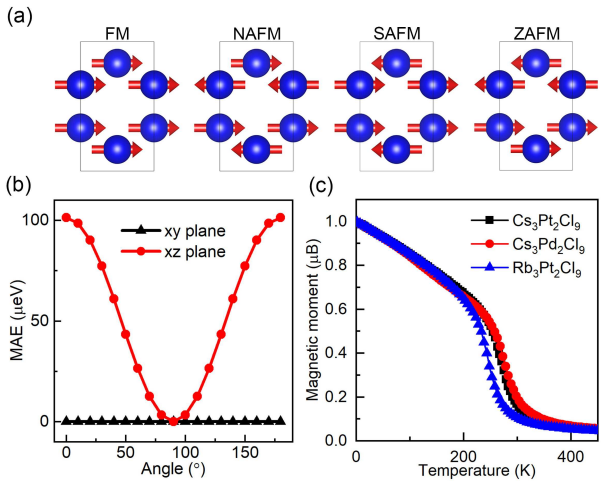}
  \caption{(a) Four possible magnetic configurations: FM, NAFM, SAFM and ZAFM. (b) The magnetic anisotropic energy (MAE) of monolayer Cs$_3$Pt$_2$Cl$_9$,with red and black curves corresponding to the relative energy in x-z and $x$-$y$ planes respect to that along x direction. (c) The normalized magnetic moment of monolayer Rb$_3$Pt$_2$Cl$_9$, Cs$_3$Pt$_2$Cl$_9$ and Cs$_3$Pd$_2$Cl$_9$ as a function of temperature by Monte Carlo simulations.}
  \label{fig2}
\end{figure}

To determine the ground states of monolayer A$_3$B$_2$C$_9$, four representative magnetic configurations, i.e., ferromagnetic (FM), N\'eel antiferromagnetic (NAFM), stripe AFM (SAFM) and zigzag AFM (ZAFM) as shown in Fig.~\ref{fig2}(a), are considered. The GGA+U scheme is used to describe the onsite Coulomb repulsion on the \emph{d} electrons of Pd and Pt atoms~\cite{GGA-U}. The value of U is set to be 3.5/2.5 eV for Pd/Pt, consistent with the result of magnetic moment obtained from HSE06 calculation. The energies of different magnetic configurations are summarized in Table S1, indicating that FM configuration is the most stable state for all monolayer A$_3$B$_2$C$_9$. The magnetic anisotropy energy (MAE) calculations show that the magnetization prefers to lie inside the $x$-$y$ plane, such as Cs$_3$Pt$_2$Cl$_9$ shown in Fig.~\ref{fig2}(b).  The larger atomic number of occupation for each site leads to the enhanced MAE, e.g., the MAE in x-z plane increases from 62.2 $\mu$eV (Rb$_3$Pd$_2$Cl$_9$) to 2.6 meV (Cs$_3$Pt$_2$Br$_9$). Although the Mermin-Wagner theorem prohibits long-range magnetic order in two-dimensional isotropic systems, a quasi-long-range order is allowed blow the Berezinskii-Kosterlitz-Thouless critical temperature, such as VS$_2$ and FeCl$_2$~\cite{Mermin-Wagner,VS2,FeCl2}. Based on XY model, the Berezinskii-Kosterlitz-Thouless transition temperature is estimated up to the third neighbor coupling strength. The spin Hamiltonian is expressed as:

\begin{equation}
\begin{split}
H=&-\sum_{\langle i,j \rangle}J_{1}(S_{i}^xS_{j}^x+S_{i}^yS_{j}^y)-\sum_{\langle \langle i,j \rangle \rangle}J_{2}(S_{i}^xS_{j}^x+S_{i}^yS_{j}^y)\\
&-\sum_{\langle \langle \langle i,j \rangle \rangle \rangle}J_{3}(S_{i}^xS_{j}^x+S_{i}^yS_{j}^y),
\end{split}
\end{equation}
where $S^{x,y}$ is the spin operator, $\langle ...\rangle$, $\langle\langle ... \rangle\rangle$ and $\langle\langle \langle ... \rangle\rangle\rangle$ denote the sum over the nearest, the second and the third neighboring sites. $J_{1,2,3}$ denote the nearest, the second and the third neighbor exchange interaction strengths, which can be extracted from the following equations:
\begin{equation}
\begin{aligned}
E_{\rm{FM/NAFM}} = E_{0}+(\pm 3J_{1}+6J_{2} \pm 3J_{3})\left|\overrightarrow{S}\right|^2\\
E_{\rm{SAFM/ZAFM}} = E_{0}+(\mp J_{1}-2J_{2}\pm 3J_{3})\left|\overrightarrow{S}\right|^2
\end{aligned}
\end{equation}
where $E_0$ is the ground state energy, and $S=1/2$ in our system. The exchange coupling constants of monolayer A$_3$B$_2$C$_9$ are listed in Table~I. The first neighbor exchange interactions are always negative and excess the second and the third neighbor exchange interactions obviously in favor of the FM state. By performing the Monte Carlo calculations, one can acquire the Berezinskii-Kosterlitz-Thouless temperature. The critical temperatures of monolayer A$_3$B$_2$C$_9$ range from 47 to 278K [see Fig.~\ref{fig2}(c), Fig. S7 and Fig. S8], indicating that the quasi-long-range can be observed at relatively high temperatures as that of CoPS$_3$~\cite{CoPS3}.

\begin{figure}
  \centering
  \includegraphics[width=0.45\textwidth]{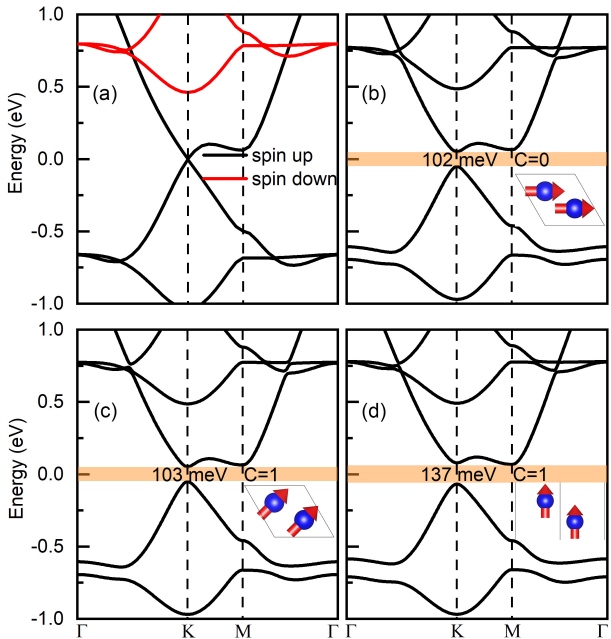}
  \caption{(a) Band structure of monolayer Cs$_3$Pt$_2$Cl$_9$ without spin-orbit coupling. (b)-(d) Band structures with spin-orbit coupling when the magnetization direction along $\phi = 0^{\circ},~45^{\circ}$ and $z$ direction, respectively.}
  \label{fig3}
\end{figure}

\begin{figure}
  \centering
  \includegraphics[width=0.45\textwidth]{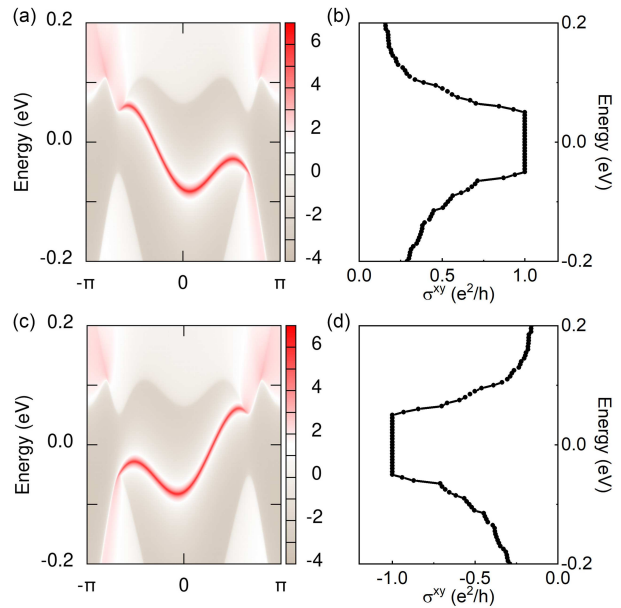}
  \caption{Energy spectra of semi-infinite ribbon of monoalyer Cs$_3$Pt$_2$Cl$_9$ with the in-plane magnetization along $\phi = 45^{\circ}$ (a) and $-45^{\circ}$ (c). The corresponding anomalous hall conductivity with $\phi = 45^{\circ}$ (b) and $-45^{\circ}$ (d).}
  \label{fig4}
\end{figure}

\textit{Band Structures and Topological Properties---.} Now, let us move to the electronic properties of monolayer A$_3$B$_2$C$_9$. As depicted in Fig.~S4, all A$_3$B$_2$C$_9$ monolayers are half-metals with the Dirac point being located at K point. And the energy range with only one spin is over 1.0 eV, facilitating to build spintronic devices with various materials. At the high symmetric point K, the gapless Dirac point is protected by the group of wave vector, i.e., D${_3^2}$ space group. Along the high symmetry K-M line, the lowest unoccupied band crosses the Fermi level in five monolayer A$_3$B$_2$C$_9$, resulting in the absence of  global gaps in the presence of spin-orbit coupling.
After taking spin-orbit coupling into consideration, due to the lack of vertical mirror symmetry at K, a band gap opens regardless of the magnetization direction. Three Cl-based monoalyers, i.e., Cs$_3$Pd$_2$Cl$_9$, Cs$_3$Pt$_2$Cl$_9$ and Rb$_3$Pt$_2$Cl$_9$, exhibit global band gaps. Hereinbelow, we take the monoalyer Cs$_3$Pt$_2$Cl$_9$ as an example.  Without spin-orbit coupling, the band of Cs$_3$Pt$_2$Cl$_9$ displays half-metallic Dirac point as shown in Fig.~\ref{fig3}(a). When spin-orbit coupling is considered, a direct gap of 102 meV opens at the Dirac point as demonstrated in Fig.~\ref{fig3}(b). The topological properties of the gap depend on the magnetization direction, i.e., (i) when the magnetization direction, e.g., along $x$ axis, is perpendicular to the vertical mirror plane, the band gap would be trivial since the symmetry restriction~\cite{QAHE-inplane1,QAHE-inplane2}; (ii) the remaining magnetization direction breaks the vertical mirror symmetry and induces a topologically nontrivial gap harboring QAHE. When the magnetization is along $\phi = 45^{\circ}$, as displayed in Fig.~\ref{fig3}(c), a topologically nontrivial gap with Chern number of $\mathcal{C}=1$ can be produced. An out-of-plane magnetization further results in a larger topologically nontrivial gap up to 137 meV as shown in Fig.~\ref{fig3}(d). Moreover, the QAHE is magnetization-orientation-dependent with alternating Chern numbers of $\mathcal{C}=\pm1$~\cite{LaCl}. As displayed in Fig.~\ref{fig4}, the Chern number can be tuned from +1(-1) to -1(+1) by rotating the magnetization direction in $x$-$y$ plane. Due to the negligible MAE in $x$-$y$ plane, a tiny magnetic field can be used to rotate the in-plane magnetization direction and enhance the effective anisotropy as employed in Cr$_2$Ge$_2$Te$_6$ film~\cite{Cr2GeTe6}, indicating the feasibility of magnetization direction tunable topological phases in our proposed systems.

\textit{Materials Synthesis---.} As a big materials family, halide perovskite consists of a number of two-dimensional layered materials. Except the 111-type In-based halide perovskite Cs$_3$In$_2$Cl$_9$, it was found that Cs$_3$Fe$_2$Cl$_9$ has been synthesized in 1954~\cite{Cs3Fe2Cl9}. This means that through proper modulation of synthesis, it is promising to produce the materials of monolayer A$_3$B$_2$C$_9$ for the study of high temperature QAHE.

\textit{Summary---.} We theoretically proposed a new two-dimensional halide perovskite family A$_3$B$_2$C$_9$ (A=Rb, Cs; B=Pd, Pt; C=Cl, Br). We find that five of them are half-metals with wide fully spin-polarized energy range. When spin-orbit coupling is further considered, we show that monoalyer Cs$_3$Pd$_2$Cl$_9$, Cs$_3$Pt$_2$Cl$_9$ and Rb$_3$Pt$_2$Cl$_9$ can open up topologically nontrivial band gaps of $\sim$ 100 meV to realize high temperature QAHE with Chern numbers of $\mathcal{C}=\pm1$. A small external magnetic field can be applied to tune the sign of Chern number by altering the magnetization direction. Our proposed material systems of A$_3$B$_2$C$_9$ are experimentally feasible, because of the successful synthesis of Cs$_3$In$_2$Cl$_9$ and Cs$_3$Fe$_2$Cl$_9$ crystal. Our findings provide a new platform for realizing high temperature QAHE in the intrinsic two-dimensional halide perovskite.

\textit{Acknowledgements---.} This work was financially supported by the NNSFC (No. 11974327), Fundamental Research Funds for the Central Universities (WK3510000010, WK2030020032), Anhui Initiative in Quantum Information Technologies. We also thank the supercomputing service of AM-HPC and the Supercomputing Center of University of Scienceand Technology of China for providing the high performance computing resources.

\end{document}